\documentclass[cits]{PoS}

\title{Magnetic fields in star-forming galaxies at high and low redshift}

\ShortTitle{Magnetic fields in star-forming galaxies at high and low redshift}

\author{\speaker{Timothy Garn}, Dominic Ford, Paul Alexander, David A.\
        Green, Julia M.\ Riley\\
        Astrophysics Group, Cavendish Laboratory, 19 J.~J.~Thomson
        Ave., Cambridge CB3~0HE, U.K.\\
        E-mail: \email{tsg25@cam.ac.uk}}

\usepackage[hang,raggedright]{subfigure}
\graphicspath{{./Images/}}
\abstract{As part of an ongoing series of deep GMRT surveys we have
          observed the Spitzer extragalactic First Look Survey field,
          producing the deepest wide-field 610-MHz survey published to
          date.  We reach an rms noise level of 30~$\mu$Jy~beam$^{-1}$
          before primary beam correction, with a resolution of
          $\sim$6~arcsec over an area of $\sim$4~deg$^{2}$.  By
          combining these observations with the existing 1.4-GHz VLA
          survey produced by Condon et al.\ (2003), along with
          infrared data in up to seven wavebands from the Spitzer
          Space Telescope, optical photometry from SDSS and a range of
          spectroscopic redshift surveys, we are able to study the
          relationship between radio luminosity and star formation
          rate in star-forming galaxies up to $z \sim 1$.  The large
          amount of multi-wavelength data available allows
          $k$-corrections to be performed in the radio due to the
          knowledge of the radio spectral index, and in the infrared
          through the use of a semi-empirical radiative transfer model
          which models star-forming regions, warm dust surrounding
          these regions, and diffuse interstellar dust, taking into
          account the star formation rate, star formation history and
          hydrogen column density within each galaxy.  A strong
          correlation is seen between radio luminosity and the
          infrared-derived star formation rates, which is best fit by
          a slightly non-linear power-law.  We look for cosmic
          evolution in the comparative radio brightness of
          star-forming galaxies by searching for deviations away from
          the global relationship.  Any such deviation would indicate
          a systematic variation in one or more of the properties
          controlling synchrotron radiation, in particular an increase
          in the magnetic field strengths of star-forming galaxies
          over time.  The data shows no evidence for such an effect,
          suggesting that there has been little evolution in the
          magnetic fields of galaxies since $z\sim1$.}

\FullConference{From Planets to Dark Energy: the Modern Radio Universe\\
		 October 1-5 2007\\
		 The University of Manchester, UK}

\begin{document}
%----------------------------------------------------------------------
\section{GMRT surveys}
\begin{figure}
  \begin{center}
  \includegraphics[width=\textwidth]{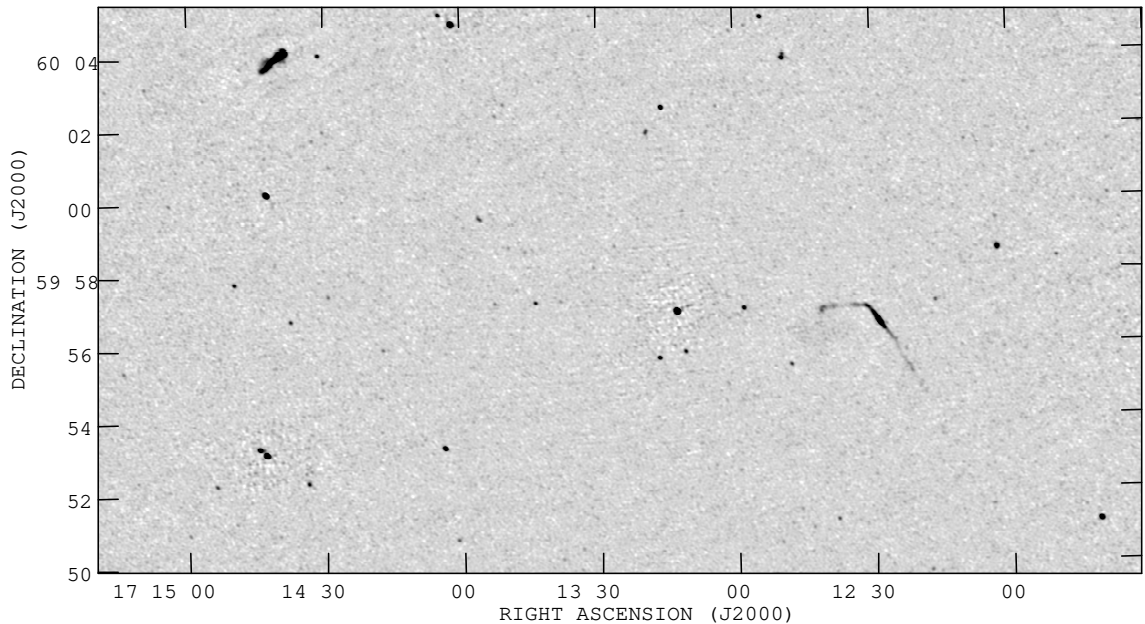}
  \caption{A sample section of the GMRT 610-MHz image of the Spitzer
  extragalactic First Look Survey field\cite{Garn07}.  The grey-scale
  ranges between $-0.2$ and 1~mJy~beam$^{-1}$.}
  \label{fig:xFLSimage}
  \end{center}
\end{figure}

We are carrying out a series of deep radio surveys with the Giant
Metrewave Radio Telescope (GMRT) at 610~MHz, targeting regions with
large amounts of existing multi-wavelength data.  The first of these,
the Spitzer extragalactic First Look Survey (xFLS) field\cite{Garn07}
covers $\sim$4~deg$^{2}$ with a resolution of $\sim$6~arcsec and rms
noise level before primary beam correction of 30~$\mu$Jy~beam$^{-1}$.
Fig.~\ref{fig:xFLSimage} shows a portion of the survey region, in
order to demonstrate the image quality.  This is the deepest
wide-field GMRT survey published to date, with 3944 sources detected
above a threshold of $5.25\sigma$.  Two other surveys have been
completed, one covering $\sim$9~deg$^{2}$ of the ELAIS-N1
region\cite{Garn08EN1} to a noise level of 70~$\mu$Jy~beam$^{-1}$,
including a deeper region with 40~$\mu$Jy~beam$^{-1}$ rms, and another
covering $\sim$5~deg$^{2}$ of the Lockman Hole\cite{Garn08LH} to a
noise level of 60~$\mu$Jy~beam$^{-1}$.  A survey of the ELAIS-N2
region is in progress, and will be completed in 2008\footnote{610-MHz
images and source catalogues can be accessed via {\href
{http://www.mrao.cam.ac.uk/surveys}{\tt
http://www.mrao.cam.ac.uk/surveys}}.}.

Our survey of the xFLS field was designed to complement the existing
VLA 1.4-GHz survey\cite{Condon03}, which has $\sim$5~arcsec resolution
and covers approximately the same area with a noise level of
23~$\mu$Jy~beam$^{-1}$.  There is a great deal of complementary data
available on the region -- in particular, this was the location of the
first of the deep surveys carried out by the Spitzer Space Telescope.
We constructed a sample of 235 star-forming galaxies within the xFLS
field, requiring all sources to have radio detections at 610~MHz and
1.4~GHz, infrared detections at 24 and 70~$\mu$m and a spectroscopic
redshift.  Where available, optical photometry from SDSS and further
infrared data from the IRAC and MIPS instruments were included in the
data set.  More details on the sample selection procedure and the
minimisation of potential sample biases can be found in
\cite{Garn08IR}.

%----------------------------------------------------------------------
\section{Radio and infrared $k$-corrections}
\begin{figure}
  \centerline{\subfigure[Radio $k$-corrections.]{
                \includegraphics[width=.5\textwidth]{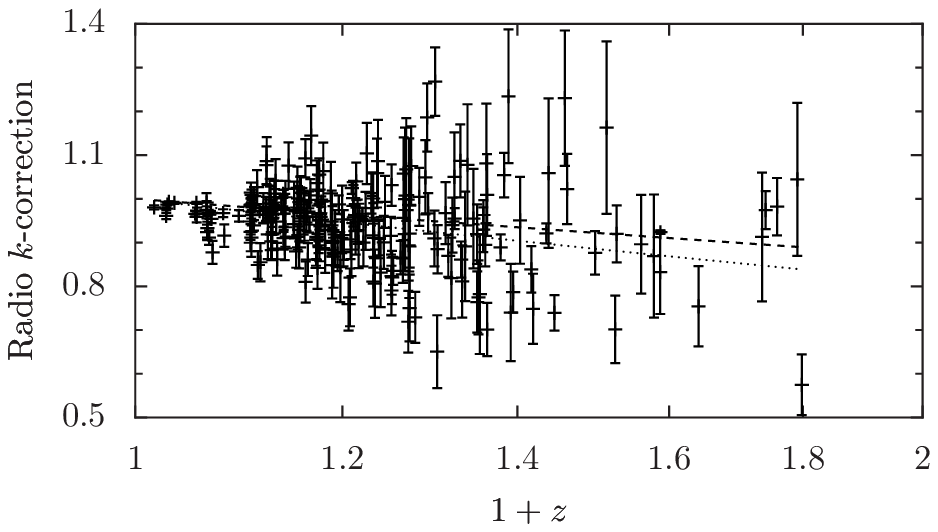}}
                \subfigure[Infrared $k$-corrections.]{
                \includegraphics[width=.5\textwidth]{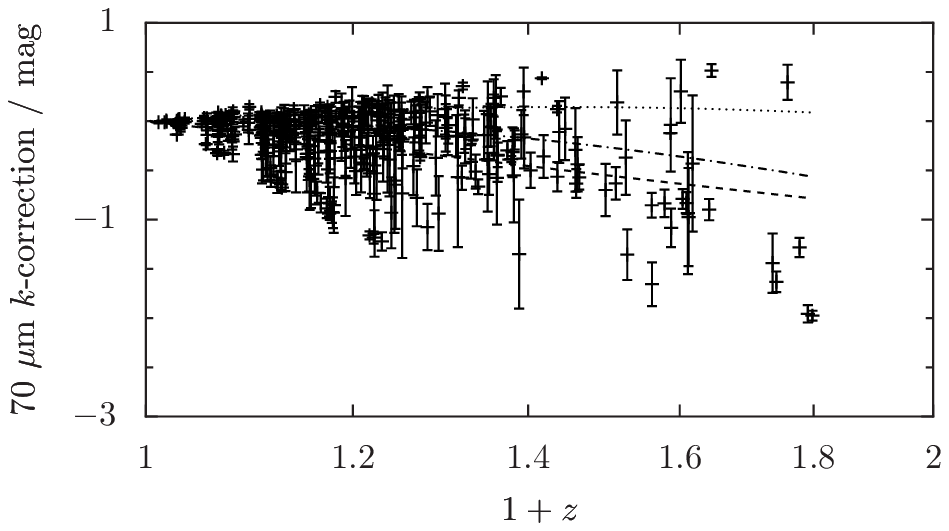}}
                }
  \caption{Radio $k$-corrections are shown in (a), calculated from the
  measured spectral index (points with error bars), $\alpha=0.7$
  (dashed line) and $\alpha=0.8$ (dotted line).  Infrared corrections
  are shown in (b), taken from the model spectra (points with error
  bars), M51 (dashed line), M82 (dotted) and Arp~220 (dash-dot).}
  \label{fig:kcorr}
\end{figure}
In order to study the luminosity of galaxies at high redshift it is
necessary to $k$-correct the observed flux density $S_{\nu}$ at
frequency $\nu$ to give a rest-frame luminosity $L_{\nu}$ at the same
frequency.  The dominant emission mechanism in the radio frequency
range we are considering is synchrotron emission, which we assume to
follow a power-law such that $S_{\nu} = S_{0} \nu^{-\alpha}$, and the
luminosity of a source at redshift $z$ and luminosity-distance $d_{\rm
L}$ is therefore given by $L_{\nu} = 4 \pi d_{\rm L}^2 (1 + z)^{\alpha
- 1} S_{\nu}$.  We assume a flat cosmology with $\Omega_{\Lambda} =
0.74$ and $H_{0} = 72$~km~s$^{-1}$~Mpc$^{-1}$.  Since radio
observations were previously available at only one frequency, radio
$k$-corrections carried out in the xFLS region by other authors have
assumed a single spectral index for all sources (either 0.7 or 0.8).
Using the two frequency data, we construct a spectral index
distribution for our sources\cite{Garn08IR} and find a peak near 0.8,
but considerable spread.  Fig.~\ref{fig:kcorr}a shows our calculated
radio $k$-corrections compared with those using a constant value for
$\alpha$ of 0.7 or 0.8; using an assumed spectral index leads to a
significant difference in $k$-correction for individual sources.  In
particular, the use of the observed spectral index permits
$k$-corrections which are greater than unity.

Model optical-infrared spectral energy distributions (SEDs) were
generated using the Starburst99 stellar spectral synthesis
code\cite{Leitherer99}, together with a semi-empirical radiative
transfer model for the propagation of radiation through the dusty
interstellar medium\cite{Garn08IR,Ford08}.  The model is made up of
star-forming regions, warm dust associated with these regions, and a
diffuse distribution of cold dust throughout the galaxy.  The
star-formation rate (SFR), star formation history and dust mass were
varied in order to create a grid of 2145 galaxy models, which were
weighted according to their $\chi^{2}$ misfit to each source.  From
these, we calculate weighted $k$-corrections and SFRs for each galaxy.
Previous works have typically taken local sources such as M82 to be
representative of the whole star-forming galaxy population -- however,
the Malmquist bias dictates that the average luminosity of survey
sources will increase with redshift, potentially making this
assumption invalid.  In Fig.~\ref{fig:kcorr}b we show the 70-$\mu$m
$k$-correction for all sources in our sample, along with those derived
from three well-studied local sources (M51, M82 and Arp~220).  The
$k$-corrections derived from our model fitting follow the same general
trends as for the local sources, but again the difference between
individual galaxies and the local sources increases at higher
redshift, by up to $\sim$1~mag.

%----------------------------------------------------------------------
\section{The relationship between star formation rate and radio
  luminosity} 
\begin{figure}
  \centerline{\subfigure[1.4-GHz luminosity against infrared SFR --
                the best-fit line is a power law given by
                Equation~{\protect \ref{eq:LSFR}}.]{
                \includegraphics[width=.5\textwidth]{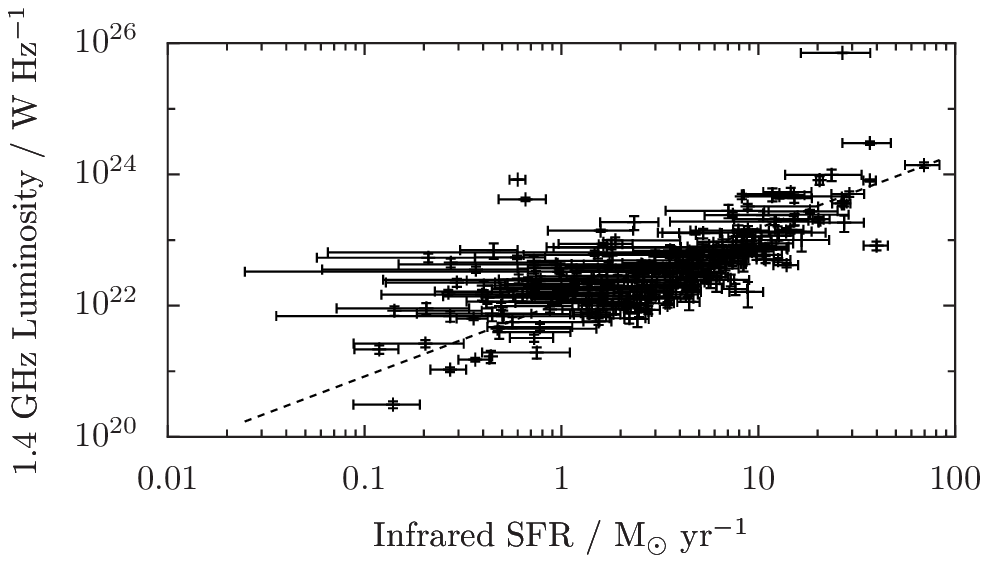}} 
              \subfigure[The redshift dependence of the relationship
                seen in (a).  The solid line represents the global
                value from Equation~{\protect \ref{eq:LSFR}}.]{
                \includegraphics[width=0.5\textwidth]{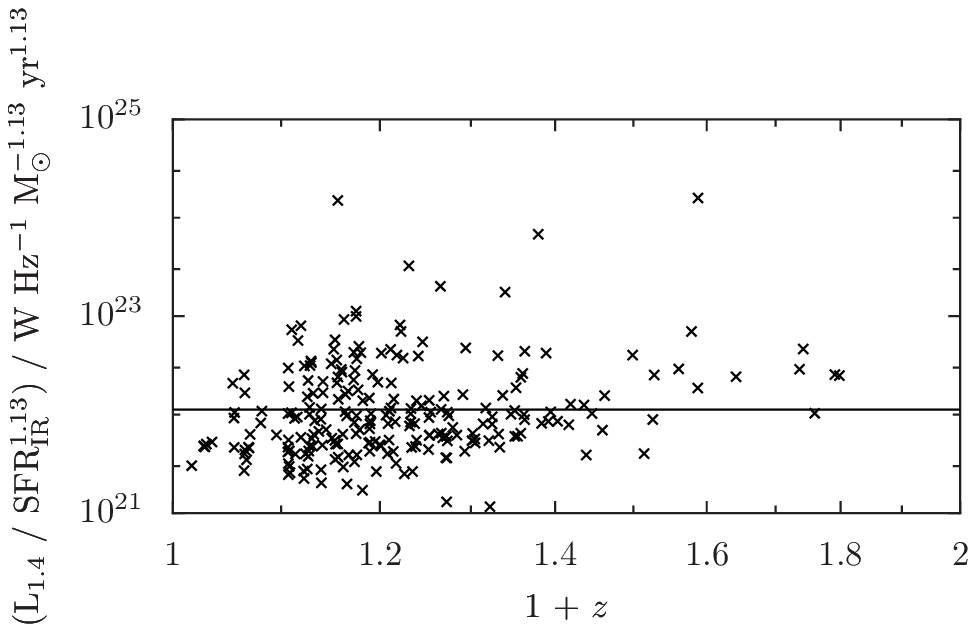}}}
  \caption{Comparison between radio luminosity and SFR (a), and the
  variation in this relationship with redshift (b).}
  \label{fig:SFR}
\end{figure}

The radio luminosity of a galaxy can be related to its current
supernova rate, and therefore to its SFR about 30~Myr earlier (the
approx.\ lifetime of the least massive stars to form a supernova)
through the use of an assumed Initial Mass Function.  There are large
uncertainties present in the normalisation between supernova rate and
luminosity, and there is no a-priori reason to assume that the complex
processes involved in synchrotron emission should lead to a linear
relationship between supernova rate and radio luminosity.  We
therefore look directly at the relationship seen between 1.4~GHz
luminosity $L_{1.4}$ and SFR, seen in Fig.~\ref{fig:SFR}a, and find
that the relationship is best fit by a power-law given by
\begin{equation}
{\rm log}_{10}(L_{1.4}) = (22.05\pm0.04) + (1.13\pm0.07)\times{\rm
  log}_{10}({\rm SFR}).
\label{eq:LSFR}
\end{equation}
There are a few sources that are significantly more radio-bright than
this relationship -- these are likely to have some radio emission
resulting from AGN activity as well as due to supernovae, since the
AGN rejection in \cite{Garn08IR} was performed based upon optical
spectroscopy alone.  The quoted results remain consistent when these
sources are removed from the analysis.

We probe the redshift dependence of the relationship between radio
luminosity and SFR, by looking at deviations away from the global form
found in Equation~\ref{eq:LSFR}.  This allows us to search for cosmic
evolution in the comparative radio brightness of star-forming
galaxies.  Fig.~\ref{fig:SFR}b shows the redshift dependence of
$L_{1.4}$/SFR$^{1.13}$ -- no significant variation is seen out to a
redshift of 1.

\section{Discussion \& Conclusions}
The large amount of multi-wavelength information available on regions
such as the Spitzer extragalactic First Look Survey field mean that it
is possible to create large samples of star-forming galaxies which
contain a significant amount of spectral information.  This allows
individual $k$-corrections to be performed, using the knowledge of the
optical stellar emission, infrared thermal dust emission and radio
synchrotron emission from each galaxy in the sample.

We find the relationship between radio luminosity and star formation
rate has a slightly non-linear form with power law index of
$1.13\pm0.07$.  By comparing the radio luminosity and infrared-derived
star formation rates of galaxies between $z \sim 1$ and the present
day, we find no evidence for significant evolution in the radio
luminosity of star-forming galaxies.  

Any systematic variation over time in the radio luminosity of galaxies
with particular SFR would imply an evolution in one or more of the
properties that controls synchrotron radiation, and lead to the tight
infrared / radio correlation\cite{Appleton04}.  In particular, an
increase in the magnetic field strength of galaxies since redshift 1
($\sim$7.7~Gyr) would lead to greater synchrotron emission from the
electron population around supernova remnants and a greater chance of
electron `confinement' -- electrons radiating away all of their energy
within a galaxy, rather than escaping.  This would lead to an increase
in the radio luminosity for sources with given SFR at lower redshifts.
No such effect is seen, suggesting that there has been little
evolution in the magnetic fields of galaxies over this redshift range.

\acknowledgments 
\noindent {\small We thank the staff of the GMRT who have made these
observations possible.  The GMRT is operated by the National Centre
for Radio Astrophysics of the Tata Institute of Fundamental Research,
India.  TG and DCF both thank the UK STFC for Studentships.  This
work has made use of the distributed computation grid of the
University of Cambridge (\textsc{CamGRID}).}
%----------------------------------------------------------------------

\end{document}